\documentclass[reprint,
 amsmath,amssymb,
 aps,floatfix,twocolumn,floatfix]{revtex4-2}
\makeatletter
\def\endwidetext{%
  \par
  \vskip1.5\baselineskip
  \twocolumngrid
  \global\@ignoretrue
}
\makeatother
\usepackage{graphicx}
\usepackage{dcolumn}
\usepackage{bm}
\usepackage{amsmath}
\usepackage{braket}
\usepackage{xcolor}
\usepackage[T1]{fontenc}
\usepackage{comment}
\usepackage[export]{adjustbox}
\usepackage{array}
\usepackage{multirow}
\usepackage{booktabs}
\usepackage{float}
\usepackage{makecell}
\usepackage{amssymb}
\usepackage{appendix}
\usepackage{accents}

\begin{document}

\preprint{APS/123-QED}
\title{Localized to delocalized spatial quantum correlation evolution in structured bright
twin beams}
\author{Jerin A Thachil}
 \author{Chirang R. Patel}
 \author{U. Ashwin}
\author{Ashok Kumar}
 \email{ashokkumar@iist.ac.in}
\affiliation{Department of Physics, Indian Institute of Space Science and Technology\\
 Thiruvananthapuram, Kerala, 695547, India
}
\date{\today}

\begin{abstract}

Quantum correlations in the spatial domain hold great promise for applications in quantum imaging, quantum cryptography and quantum information processing, owing to the infinite dimensionality of the associated Hilbert space. Here, we present a theoretical investigation, complemented by experimental measurements, of the propagation dynamics of the spatial quantum correlations in bright structured twin beams generated via a four-wave mixing process in a double-$\Lambda$ configuration in atomic vapor. We derive an analytical expression describing the evolution of the spatial quantum correlation distribution from the near field to the far field. To qualitatively support the theoretical predictions, we perform experiments measuring intensity-difference noise between different spatial subregions of the twin beams as they propagate from the near field to the far field. The presence of quantum correlations is manifested as squeezing in the intensity difference noise measurement. With a Gaussian pump, we observe localized correlations in the near field and localized anti-correlations in the far field. In contrast, with a structured Laguerre-Gaussian pump, there is a transition from localized correlations in the near field to delocalized correlations in the far field. The present results offer valuable insights into the fundamental behavior of spatial quantum correlations and open possibilities for potential applications in quantum information, quantum imaging and sensing.

\end{abstract}

\maketitle
\section{\label{sec:level1}INTRODUCTION}

Spatial quantum correlations offer immense potential in quantum information applications, owing to the infinite dimensionality of their Hilbert space \cite{Kolobov, Mirh, Otte}. Such high dimensionality enables parallel information encoding, enhanced security in quantum communication protocols, and exponential speed-ups in quantum computing \cite{Lassen,Otte, Mirh, Wang}. Moreover, spatial quantum correlations extend quantum advantages beyond the time domain into the spatial domain, opening new possibilities in quantum imaging and metrology with enhanced resolution and sensitivity \cite{Brida, Treps, Dowran}.

Entangled twin beams of light serve as ideal candidates for studying and manipulating spatial quantum correlations \cite{Boyer, Boyer2}. The generation of such quantum states relies on nonlinear parametric processes like spontaneous parametric down-conversion (SPDC) or four-wave mixing (FWM), where the nonlinear optical interaction of a strong pump beam with the medium leads to the generation of photon pairs. Such processes inherently conserve both energy and momentum: energy conservation leads to temporal quantum correlations, while momentum conservation gives rise to spatial quantum correlations. Furthermore, the spatial properties of the generated photons, including the spatial profile and the spatial quantum correlations, are governed by the spatial characteristics of the input pump beam, as determined by the phase-matching condition \cite{Boucher, Walborn, Monken, Khoury, Barbosa, Swaim, Marino, Cao, Du}. Recent studies have demonstrated that structured pump beams can be used to encode information into the spatial quantum correlations of twin beams \cite{Nirala, Verniere}. Therefore, engineering the spatial quantum correlations and understanding their evolution during propagation is crucial for both fundamental quantum optics research and practical applications \cite{Brida, Treps, Dowran, Wang}.

While most previous works have explored the spatial quantum correlation distribution in the near and far fields \cite{Brambilla, Howell,Jedrki, Edgar,Bhat, Ashok1, Ashok2, Ashok3, Ashok4, Hamar}, relatively few studies have focused on the dynamic evolution of these correlations as the beams propagate from the near field to the far field \cite{Chan, Tasca, Galinis, Haderka, Abhi}. In the near field, position cross correlations are observed due to the common birth zone of the photon pair while in the far field momentum anti-correlations are observed as a consequence of the phase matching condition. However, in the intermediate regime, the spatial quantum correlations gradually blur. Most of such studies are carried out in the single photon pair regime using SPDC sources with a Gaussian pump. Nonetheless, over the past decade, studies on spatial quantum correlations in the macroscopic regime have gained significant attention due to the ease of measurement and single shot determination of quantum correlations \cite{Swaim, Nirala, Marino, Cao, Ashok1, Ashok2, Ashok3, Ashok4, Du}. Four-wave mixing in atomic ensembles provides an efficient method to generate such macroscopically, bright twin beams; owing to the strong intrinsic nonlinearity of resonant atom-light interactions \cite{McCormick, Lukin}. Unlike optical parametric oscillators based on parametric downconversion \cite{Heidmann, Mertz}, such systems do not require any optical cavity to generate bright twin beams, and therefore, can support multiple spatial modes. Towards the spatial quantum correlations in structured bright twin beams generated with the FWM process, Marino $et.~al$ \cite{Marino} has shown delocalized spatial quantum correlations in the far field using a Laguerre-Gaussian (LG) pump beam that imparts the orbital angular momentum (OAM) to the generated twin beams. Nonetheless, a comprehensive study of how these quantum correlations evolve during propagation with structured pump beams remains to be explored. 

In this work, we present a detailed theoretical analysis of the evolution of transverse spatial quantum correlations in bright structured twin beams generated via a four-wave mixing process in a double-$\Lambda$ configuration with Gaussian and Laguerre–Gaussian pumps. We derive an analytical expression describing the propagation of the spatial quantum correlation distribution in such a configuration. To qualitatively support the theoretical predictions, we carry out an experimental study by measuring intensity-difference noise across selected subregions of the beams at different propagation planes, from the near field to the far field.

The paper is organized as follows. Sec.~\ref{Theory} introduces the theoretical model of the FWM process and presents the analytical expression for the propagated two-photon probability amplitude of structured twin beams and obtain their spatial quantum correlation distribution. Sec.~\ref{Exp setup} presents the experimental investigation and analysis, supporting theoretical results. Sec.~\ref{Conclusion} gives the concluding remarks. We present the details of theoretical calculations in an Appendix.

\section{Theory} \label{Theory}

\begin{figure}[t]
\begin{center}
    \includegraphics[width=0.38\textwidth]{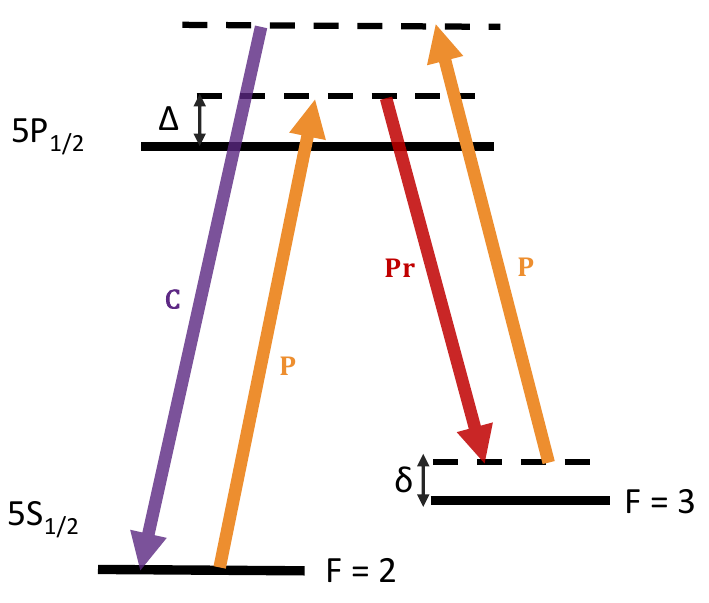} \caption{Energy levels of $^{85}$Rb involved in four-wave mixing in double-$\Lambda$ configuration.}
    \label{fig:Fig1}
\end{center}
\end{figure}

To investigate the propagation dynamics of spatial quantum correlations, we derive an analytical expression for the two-photon probability distribution for the photon pairs emitted in a FWM process in $^{85}$Rb atom, as shown in Fig.~\ref{fig:Fig1}. In this process, two pump photons (P) are absorbed by the atoms and two new photons named as probe (Pr) and conjugate (C) are generated. We examine the dependence of the two-photon probability distribution on the spatial profile of the pump beam as it evolves during the propagation. We start with the interaction Hamiltonian for the FWM process as,
\begin{align}
    \hat{H}_I = i\hbar \int d\bm{r} \chi^{(3)} \hat{E}_p^{(+)}(r,t) \hat{E}_p^{(+)}(r,t) \hat{E}_{pr}^{(-)}(r,t) \nonumber \\ \hat{E}_{c}^{(-)}(r,t) + h.c.,
\end{align}
here $\chi^{(3)}$ is the third-order nonlinear susceptibility of the medium which is assumed to be spatially independent. This holds for the FWM in hot Rb vapor cell where the atomic density is uniform across the interaction region. $\hat{E}^{(+)}(\hat{E}^{(-)})$ are the positive (negative) frequency parts of the electric field operator. The indices $p$, $pr$, and $c$ denote pump, probe, and conjugate fields, respectively. 

We consider the undepleted pump approximation and treat the pump as a classical field. The pump field propagating in the z-direction can be written as,
\begin{align}
    \hat{E}_p^{(+)}(r,t) = E(\rho, z) e^{i(k_p z - \omega_p t)},
\end{align}
where $k_p$ denotes the z-component of the pump wavevector, $\rho$ is the transverse position at the center of the nonlinear medium, and $E(\rho, z)$ is a complex function representing the slowly varying envelope of the pump’s electric field. Assuming the transverse spatial distribution of the pump remains unchanged throughout the length of the nonlinear medium, the slowly varying field amplitude is assumed to be independent of $z$, that is $E(\rho, z)$ = $E(\rho)$.

The electric field operators for the probe and conjugate can be written in the momentum basis as,
\begin{align}
   \hat{E}_{j}^{(-)}(r,t) = \left(\frac{1}{2\pi}\right)^{3/2} \int d\bm{k}_j e^{-i(\bm{k}_{j}.\bm{r} - \omega_j t)} \hat{a}_j^{\dag},
\end{align}
where the subscript $j$ = ($pr$, $c$) and $\hat{a}_j^{\dag}$ is the creation operator for the $j^{th}$ mode photon.

On substituting Eq.~(2) and (3) in Eq.~(1), and after simplifying, the Hamiltonian becomes \cite{Nirala},
\begin{align}
    \hat{H}_I = i\hbar \Gamma & \int \int d\bm{k}_{pr} d\bm{k}_c \mathcal{E}(\bm{q}_{pr} + \bm{q}_c )  \nonumber \\ &  \mbox{sinc}\left(\left[\frac{|\bm{q}_{pr}-\bm{q}_c|^2}{4k_p}-\phi\right]\frac{L}{2}\right) \hat{a}_{pr}^{\dag} \hat{a}_c^{\dag} + h.c.,
\end{align}
where $\Gamma$ = $\chi^{(3)}/(2\pi)^3$, $q_{pr}$ and $q_c$ are the probe and conjugate wavevectors in the transverse plane, and $L$ is the length of the nonlinear medium. $\phi$ = $k_p$sin$^2\theta$, where $\theta$ is the angle between pump and probe, we neglect $\phi$ term considering the small angle approximation. The sinc term accounts for the longitudinal phase-matching condition. $\mathcal{E}(q_{pr}~+~q_c)$ contains the information of pump spatial profile which is given as,
\begin{align}
    \mathcal{E}(\bm{q}_{pr} + \bm{q}_c) & = \int d\bm{\rho} \frac{1}{2 \pi} E^2_0(\bm{\rho})e^{-i(\bm{q}_{pr} + \bm{q}_c)\cdot \bm{\rho}}. 
\end{align}

The twin beam state wavefunction produced by an FWM source can be written by using first-order perturbation theory as,
\begin{align}
| \Psi_{TB} \rangle \approx &\left(1 - \frac{i\hat{H}t}{\hbar}\right) |\Psi_{0} \rangle \nonumber \\ = & ~|\Psi_{0} \rangle + A \int\int d\bm{k}_{pr}d\bm{k}_c  \mathcal{E}(\bm{q}_{pr}+\bm{q}_c) \nonumber \\ & \times \mbox{sinc}\left(\left[\frac{|\bm{q}_{pr}-\bm{q}_c|^2}{4k_p}\right]\frac{L}{2}\right)   a_{q_{pr}}^\dag a_{q_c}^\dag |\Psi_{0} \rangle.
\end{align}
 Here $A$ = $tL\Gamma$, and $| \Psi_{0} \rangle$ is the multimode vacuum state that we ignore in the further calculations.
 
 We define the two-photon probability amplitude function $\mathcal{F}$ as,
\begin{align}
    \mathcal{F} =  \mathcal{E}(\bm{q}_{pr}+\bm{q}_c) \mbox{sinc}\left(\left[\frac{|\bm{q}_{pr}-\bm{q}_c|^2}{4k_p}\right]\frac{L}{2}\right),
\end{align}
which dictates the spatial quantum correlations between the probe and conjugate photons. Note that the spatial variables $q_{pr}$ and $q_c$ appear as the sum and difference in the arguments of the pump amplitude and phase matching functions, respectively. This makes the above expression inseparable in the probe and conjugate coordinates, indicating spatial entanglement between the two modes. That is, knowledge of the momentum (or position) of one photon directly constrains that of the other. Depending on whether the pump amplitude or the phase-matching function dominates, these correlations appear as either correlations or anti-correlations. Consequently, the coupling of $q_{pr}$ and $q_c$ governs the form of the spatial correlation distribution between probe and conjugate photons.

We now consider a Laguerre-Gaussian pump with radial index $p$ = 0 and azimuthal index $l$. The electric field amplitude of the LG mode at the beam waist is given by,
\begin{align}
    E(\rho, 0) = \sqrt{\frac{2}{\pi|l|!}}\frac{1}{w}\left(\frac{\sqrt{2}\rho}{ w}\right)^{|l|} e^{\frac{- \rho ^2}{w ^2}}e^{ -il\phi},
\end{align}
where $w$ is the beam waist radius of the beam.

On substituting Eq.~(8) in Eq.~(5), we get (See Appendix),
\begin{align}
\mathcal{E}(\bm{q}_+) =  \frac{1}{\pi|l|!}\frac{w^{2|l|}}{2^{3|l| + 1}} \bm{q}_+^{2|l|} e^{\frac{- \bm{q}_+ ^2w^2}{8}}e^{ -2il\phi_+ },
\end{align}
where we have used the variable transformation as $q_+$~=~$q_{pr}$ + $q_c$,  $q_-$ = $q_{pr}$ - $q_c$ and $\phi_+$ = $\phi_{pr}$ + $\phi_c$. Thus, the two-photon amplitude function (Eq.~7) in the momentum basis representation becomes,
\begin{align}
    \mathcal{F} = \frac{1}{\pi|l|!}\frac{w^{2|l|}}{2^{3|l| + 1}} \bm{q}_+^{2|l|} e^{\frac{- \bm{q}_+ ^2 w^2}{8}}e^{ -2il\phi_+ }\mbox{sinc}\left(\frac{L|\bm{q}_-^2|}{8k_p}\right).
\end{align}

In the position basis, the two-photon amplitude function is obtained by taking the Fourier transform of Eq.~(10) as,
\begin{align}
    \mathcal{F}' = \frac{1}{(2\pi)^2}\int \int  d\bm{q_{pr}} d\bm{q_{c}} \mathcal{F} e^{-i\bm{q}_{pr}  \cdot \bm{\rho}_{pr}} e^{-i\bm{q}_{c} \cdot \bm{\rho}_{c}}. 
\end{align}
Substituting $ \mathcal{F}$ from Eq.~(10) in Eq.~(11) and further simplifying (See Appendix), we obtain,  
\begin{align}
  \mathcal{F}' =  C \int\int  dq_{-}   dq_{+} q_-  q_+^{2|l| + 1} e^{\frac{-q_+ ^2 w^2}{8}} e^{ -2il\theta_+ }  \nonumber \\ \times    \mbox{sinc}\left(\frac{L|q_-^2|}{8k_p}\right)J_{2l}(\frac{q_+\rho_+}{2})J_0(\frac{q_-\rho_-}{2}),  
\end{align}
where $C$=$\frac{1}{\pi|l|!} \frac{w ^{2|l|}}{2^{3|l| + 2}}$, and  $J_n(x)$ is the Bessel function of first kind of order $n$ with argument $x$.
\begin{figure*}[ht]
\begin{center}
    \includegraphics[width=0.97\textwidth]{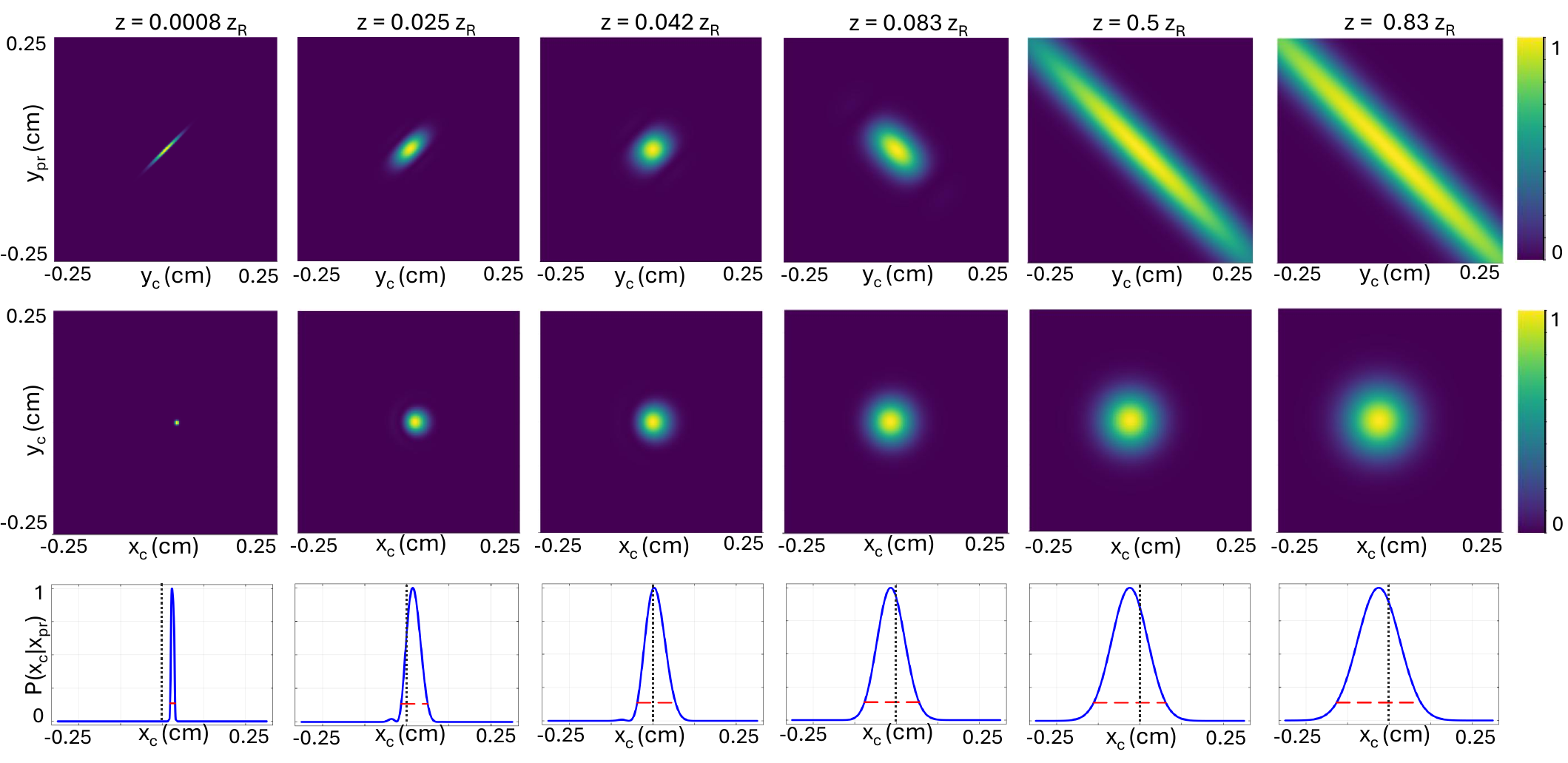} \caption{Analytical results with the Gaussian pump: (top row) two-photon probability distribution obtained by fixing $x_{pr}$ = 0.25 mm and $x_c$ = 0.25 mm, (middle row) conditional probability distribution of the conjugate with the probe coordinates fixed at $x_{pr}$ = 0.25 mm and $y_{pr}$ = 0, (bottom row) line profile obtained from the conditional probability distribution (dashed line in the line profile plot correspond to $x_c$ = 0). The red dashed lines denote the beam width defined at 1/$e^2$ of the maximum value. All propagation distances are written in terms of the pump Rayleigh range ($z{_R}$)}
    \label{fig:Fig2}
\end{center}
\end{figure*}

To obtain the two-photon probability amplitude at a given plane $z$, we apply the Fresnel propagation defined by the kernel \cite{Schneeloch, Goodman}, 
\begin{align}
    K(\rho, \rho'; z) &= \frac{k_p}{2\pi i z} e^{ik
    _pz}\exp\left[ i \frac{k_p}{2z} |\rho - \rho'|^2 \right] \nonumber \\ &= \frac{k_p}{2\pi i z}e^{ik_pz}e^{i\frac{k_p \rho^{\prime2}}{2z}}e^{i\frac{k_p\rho^{2}}{2z}}e^{-i\frac{k_p\rho^{\prime}\cdot \rho}{z}}.
\end{align}
Here we assume the magnitude of probe and conjugate wave vector are approximately equal to $k_p$. $\rho^{\prime}$ is the transverse position coordinate at plane $z$. We can propagate the two-photon amplitude function to a distance $z$ by,
\begin{align}
\tilde{\mathbb{F}} =  \int d\bm{\rho}_{pr}d\bm{\rho}_c\hspace{2pt}\mathcal{F}'\hspace{2pt} K(\rho_{pr}, \rho'_{pr}; z) \hspace{2pt} K(\rho_c, \rho'_c; z).
\end{align}

\begin{figure*}
\begin{center}
    \includegraphics[width=0.97\textwidth]{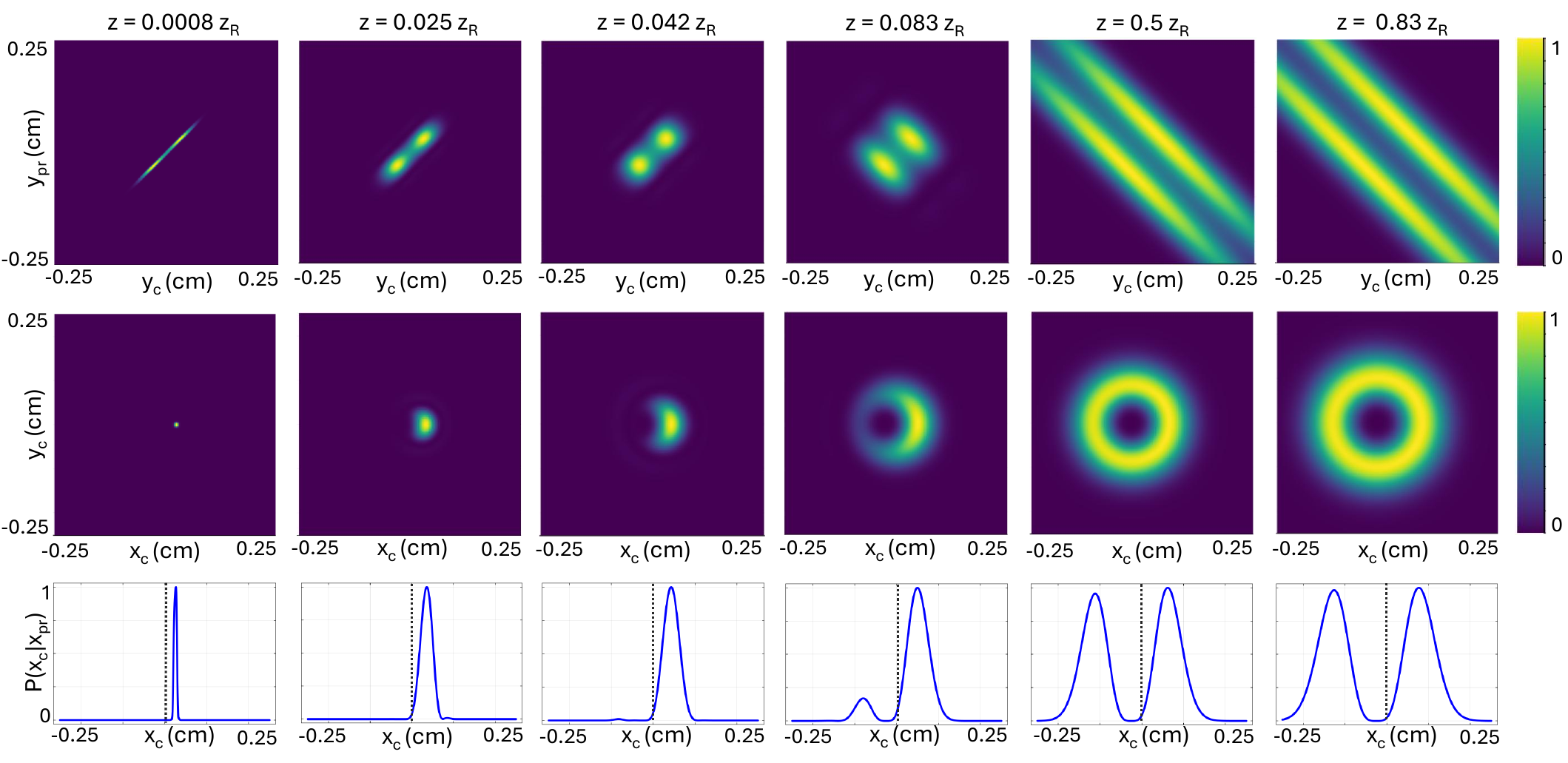} \caption{Analytical results with the Laguerre-Gaussian pump: (top row) two-photon probability distribution obtained by fixing $x_{pr}$ = 0.25 mm and $x_c$ = 0.25 mm, (middle row) conditional probability distribution of the conjugate with the probe coordinates fixed at $x_{pr}$ = 0.25 mm and $y_{pr}$ = 0, (bottom row) line profile obtained from the conditional probability distribution (dashed line in the line profile plot correspond to $x_c$ = 0). All propagation distances are written in terms of the pump Rayleigh range ($z{_R}$).}
    \label{fig:Fig3}
\end{center}
\end{figure*}

On substituting propagation kernel for probe and conjugate in Eq.~(14), we get,
\begin{align}
\tilde{\mathbb{F}} = & \frac{C}{z^2}\frac{k_p^2}{4\pi^2}e^{2ik_pz}e^{i\frac{k_p}{2z}(\rho_{pr}^{'2}+\rho_{c}^{'2})}\int \int \int\int  dq_{-}  dq_{+} d\bm{\rho}_{pr} d\bm{\rho}_{c} q_- \nonumber \\ &q_+^{2l + 1}e^{\frac{- q_+ ^2  w^2}{8}} e^{ -2il\theta_+ }  \mbox{sinc}\left(\frac{L|q_-^2|}{8k_p}\right) J_{2l}(\frac{q_+\rho_+}{2})J_0(\frac{q_-\rho_-}{2}) \nonumber \\  &e^{i\frac{k_p}{2z}(\rho_{pr}^{2}+\rho_{c}^{2})}e^{-i\frac{k_p}{z}(\rho_{pr}^{'}\cdot \rho_{pr} + \rho_{c}^{'}\cdot \rho_{c})}, 
\end{align}

We perform the variable transformation $\rho_+ = \rho_{pr} + \rho_c$ and $\rho_- = \rho_{pr} - \rho_c$. Additionally, to simplify the above equation, we use the Cosine-Gaussian approximation for the sinc function given by $\mbox{sinc}(x^2) = \cos(a x^2)\cdot e^{-b x^2}$ \cite{Bagh}. The optimization factors $a$ and $b$ are taken to be 0.39 and 0.49 for the maximum fidelity. The final expression for the propagated two-photon amplitude function $\tilde{\mathbb{F}}$ becomes (detailed calculations are given in the Appendix),
\begin{align}
    \tilde{\mathbb{F}} = \hspace{2pt}C'\alpha^{2|l|+1} (\rho^{\prime}_+)^{2|l|} &e^{ -2il\theta^{'}_+ }e^{2ik_pz} e^{-\frac{\alpha\rho^{\prime2}_+}{16}} \nonumber \\ & \times \left[ \frac{e^{\frac{-\rho _{-}^{\prime2}}{16\left(\beta + i\gamma_+\right)}}}{\left(4\beta + i\gamma_+\right)}  + \frac{e^{\frac{-\rho _{-}^{\prime2}}{16\left(\beta + i\gamma_-\right)}}}{\left(4\beta + i\gamma_-\right)}\right],
\end{align}
where $C' = \frac{1}{\pi |l|!}\frac{w^{2|l|}}{2^{5|l|+1}}$ and,
\begin{align}
\alpha =  \frac{1}{(\frac{w^2}{8}+\frac{iz}{4k_p})}, &&
\beta = \frac{bL}{8k_p}, && \gamma_{\pm} = \frac{z}{4k
_p}\pm \frac{aL}{8k_p}. \nonumber
\end{align}

In what follows, we investigate the propagation dynamics of the spatial quantum correlation distribution between the probe and conjugate beams for two cases - (a) when the pump is Gaussian, and (b) when the pump is LG. The corresponding theoretical plots are given in Figs.~\ref{fig:Fig2} and \ref{fig:Fig3} for the Gaussian and LG pumps, respectively.  In each figure, the top row corresponds to the two-photon position probability distribution, the middle row corresponds to the conditional position probability distribution of the conjugate with the probe coordinates fixed (correlation pattern at the conjugate plane), and the bottom row shows the line profile obtained from the conditional position probability distribution. All plots are normalized to unity. For the calculations, we have considered a pump beam with a waist radius 0.55 mm and a corresponding Rayleigh range $z_R$ = 1.2 m.
 
The results in Fig.~\ref{fig:Fig2} are obtained from Eq.~(16) by setting $l$=0 (Gaussian pump) and changing the propagation distances in the units of the pump Rayleigh ranges as z~=~0.0008~$z_R$ (0.1 cm), 0.025 $z_R$ (3 cm), 0.042 $z_R$ (5 cm), 0.083 $z_R$ (10 cm), 0.5 $z_R$ (60 cm) and 0.83 $z_R$ (100 cm) from the exit plane of the nonlinear medium. As it can be seen in the two-photon probability distribution (top row), we find that in the near field (z = 0.0008~$z_R$), the probe and conjugate photons have the maximum probability of arriving at the same transverse position, indicating strong position correlations in the near field. This is due to the localized emission of photon pairs at the source plane. In the intermediate region, the correlations gradually get blurred with the beam propagation. In the far field (0.83 $z_R$), they become anti-correlated in position. This is a consequence of the momentum conservation. The same dynamics can be observed in the conditional position probability distribution (middle and last row). We observe a highly localized peak in the near field (z = 0.0008~$z_R$). Here, the width (1/$e^2$ of the maximum value) of the conditional probability distribution gives a measure of the coherence area \cite{Ashok3}. As $z$ increases, the width of the peak increases monotonically. Moreover, the center of the pattern gradually shifts (see Fig.~\ref{fig:Fig2} last row) in the negative x-direction indicating the emergence of anti-correlations in the far field, vertical dashed lines in the plots are shown to mark the center of the conjugate beam ($x_c$~=~0). 

Figure \ref{fig:Fig3} presents the correlation distribution results for an LG pump with azimuthal index $l$ = 1. As with the Gaussian pump, the two-photon probability distribution exhibits strong position correlations in the near field (z = 0.0008~$z_R$).  As the photon pair propagates away from the exit plane of the cell, the correlations gradually spread out. However, unlike the Gaussian case, two distinct bright regions are observed in the correlation distribution, corresponding to the areas where probe and conjugate photon pairs are most likely to be found. This structure arises from the ring-shaped intensity profile of the LG pump, which generates photon pairs only in non-zero intensity regions of the pump. Since the pump has zero intensity at the center, no photon pairs are produced there. As the photon pairs propagate toward the far field, these bright regions broaden and stretch in the opposite directions, resulting in two parallel off-diagonal features in the distribution. The conditional probability distribution (middle row and bottom row) shows a narrow, localized peak at z = 0.0008~$z_R$, similar to the case of the Gaussian pump, implying that the near field spatial correlation pattern does not depend on the pump spatial profile. With propagation, the correlation pattern starts to expand unevenly and gradually forms an asymmetric ring at z$\sim$0.083 $z_R$. On further propagation, the correlation pattern eventually becomes a symmetric doughnut profile in the far field. This behaviour can be understood from Eq.~(10), which indicates that the far-field two-photon amplitude is determined by the Fourier transform of the square of pump spatial amplitude profile. Since LG beams possess a much broader angular spectrum than Gaussian beams, they produce correlations that are distributed over the entire transverse extent of the beams, i.e., they are delocalized.
 \begin{figure}[t]
\begin{center}
    \includegraphics[width=0.47\textwidth]{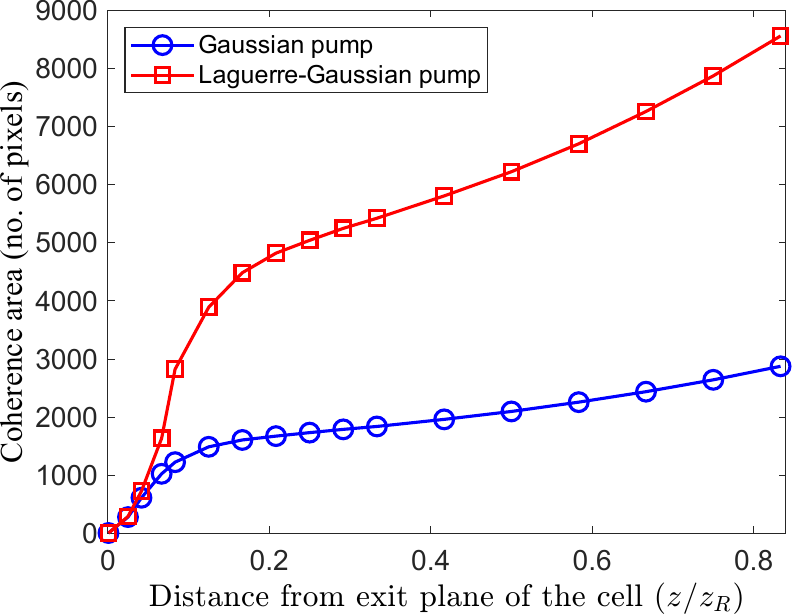} \caption{The theoretically estimated coherence areas of the twin beams plotted against the propagation distance from the cell for the Gaussian and LG pump.}
    \label{fig:Fig4}
\end{center}
\end{figure}

 \begin{figure*}[t]
  \includegraphics[width=0.8\textwidth]{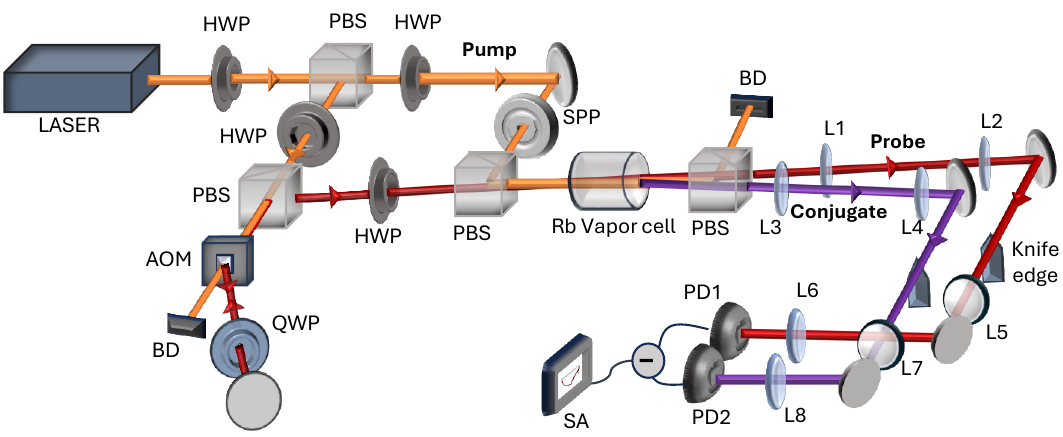}
  \caption{Schematic layout of the experimental setup used to study spatial quantum correlation dynamics of bright twin beams. SPP, Spiral phase plate; HWP, Half-wave plate; QWP, Quarter-wave plate; PBS, Polarizing beamsplitter; SA, Spectrum analyzer, L1-L8, lenses; PD1, PD2, Photodetector; BD, Beam dump.}
  \label{fig:Fig5}
\end{figure*}
Due to the asymmetry of the correlation pattern observed in the intermediate propagation regime for the LG pump, it is not trivial to define a consistent width across all propagation planes, similar to the case of the Gaussian pump. Instead, we estimate the size of the coherence area by counting the number of pixels in the correlation pattern (middle row in Fig.~\ref{fig:Fig2} and \ref{fig:Fig3}) whose values exceed 1/$e^2$ ($\approx 13.53\%$) of the maximum value. We then plot such estimated coherence areas against the propagation distance in Fig.~\ref{fig:Fig4} for the Gaussian and LG pump cases. In the near field (up to $\sim$0.042 $z_R$), the coherence area expands at nearly the same rate for both cases. In this regime, the size of the coherence area is primarily determined by the longitudinal phase-matching condition, which depends on the parameters of the nonlinear medium. Upon further propagation, the coherence area grows more rapidly for the LG pump than for the Gaussian pump. This behavior arises because, in the far field, the correlation distribution is governed by the function $\mathcal{E}(q_{pr}+q_c)$, which reflects the pump’s angular spectrum. The broader angular spectrum of the LG pump leads to a wider spread of correlations and thus a faster increase in the coherence area compared to the Gaussian case. In the intermediate regime, both pump amplitude and phase-matching function have a comparable influence on the correlation distribution. The relative influence of these two competing functions on the correlation distribution can also be inferred from analyzing the width of the two-photon probability distribution along the $\pm$45$^o$  directions.

\section{Experimental Investigation and Analysis} \label{Exp setup}

In order to investigate spatial quantum correlation dynamics experimentally, we use bright twin beams generated with a FWM process in hot $^{85}$Rb vapor. In what follows, we give a detailed description of the experimental setup and the results supporting our theoretical findings.

We implement the FWM process in a double-$\Lambda$ configuration on the D1 line ($5{}^{2}S_{1/2}$ $\rightarrow$ $5{}^{2}P_{1/2}$, 795 nm) of ${}^{85}$Rb, as shown in Fig.~\ref{fig:Fig1}. In this process, two input pump photons from a single pump beam are converted into a correlated pair of probe and conjugate photons. The pump and probe are resonant with a two-photon Raman transition between hyperfine ground levels F=2 and F=3. In the presence of a weak seed probe beam at the input, the FWM process gets stimulated and produces bright probe and conjugate beams at the output. Here, the seed beam also acts as a mode selector, fixing the probe topological charge to $l_{pr}$ and projecting the conjugate into the complementary OAM mode with charge $l_{c} = 2l_{\text{pump}} - l_{pr}$, in accordance with OAM conservation. While the resulting far-field correlation pattern depends on which detector is scanned, the size of the coherence area, and consequently the degree of delocalization, remains governed by the spatial profile of the pump.

A schematic of the experimental setup to generate bright twin beams and study their spatial quantum correlation properties is shown in Fig.~\ref{fig:Fig5}. We use a continuous wave Titanium-Sapphire laser to generate a strong pump beam ($\approx$~550~mW). A small fraction of the pump power is frequency downshifted by $\approx$ 3 GHz using an acousto-optic modulator (AOM) to generate the input seed probe. The orthogonally polarized pump and seed probe are then made to overlap at an angle of $\approx$ 0.4~$^{\circ}$ inside a 12 mm long Rb vapor cell. The cell is maintained at a temperature of 103~$^{\circ}$C. The pump and probe are focused at the center of the cell with waist diameters of 1.1 mm and 0.7 mm, respectively. After passing through the cell, the transmitted pump beam is separated from the twin beams using a polarizing beamsplitter (PBS). To convert the Gaussian input pump or seed probe beam into a LG mode, a spiral phase plate (SPP) is placed in the respective beam path before the vapor cell. 

To obtain the near field measurements, separate 4f-imaging systems are used for both probe (L1 and L2) and conjugate (L3 and L4) that image the exit plane of the vapor cell to an intermediate plane. To investigate the spatial distribution of quantum correlations between the twin beams, we select different spatial regions within the probe and conjugate beams at this intermediate plane and measure their intensity difference noise. The spatial regions within each beam are selected by clipping them with a knife edge mounted on a translational stage. The optical power of each beam is recorded before and after clipping using a power meter, and the ratio is used to calculate the fraction of intensity blocked. Another 4f-imaging system is used to image the intermediate plane onto a balanced photodetector, and the intensity difference noise between the twin beams is analyzed using a spectrum analyzer. To investigate the propagation dynamics of spatial quantum correlations, this measurement is repeated at various propagation distances from the exit plane of the cell by moving the knife edge, imaging systems and the photodetectors. For far-field measurements, the initial 4f-imaging system is replaced by a Fourier lens, and clipping is performed at the Fourier plane using knife edge. To obtain the shot noise limit (SNL), a fraction of the pump power is taken out before the cell, split equally, and sent to a balanced photodetector. 

In what follows, we study the spatial quantum correlation distribution of the generated probe and conjugate beams for three input configurations - (i) the pump and seed probe are Gaussian (ii) pump is Gaussian and seed probe is LG and (iii) pump is LG and seed probe is Gaussian.

\begin{figure}[b]
\begin{center}
    \includegraphics[width=0.46\textwidth]{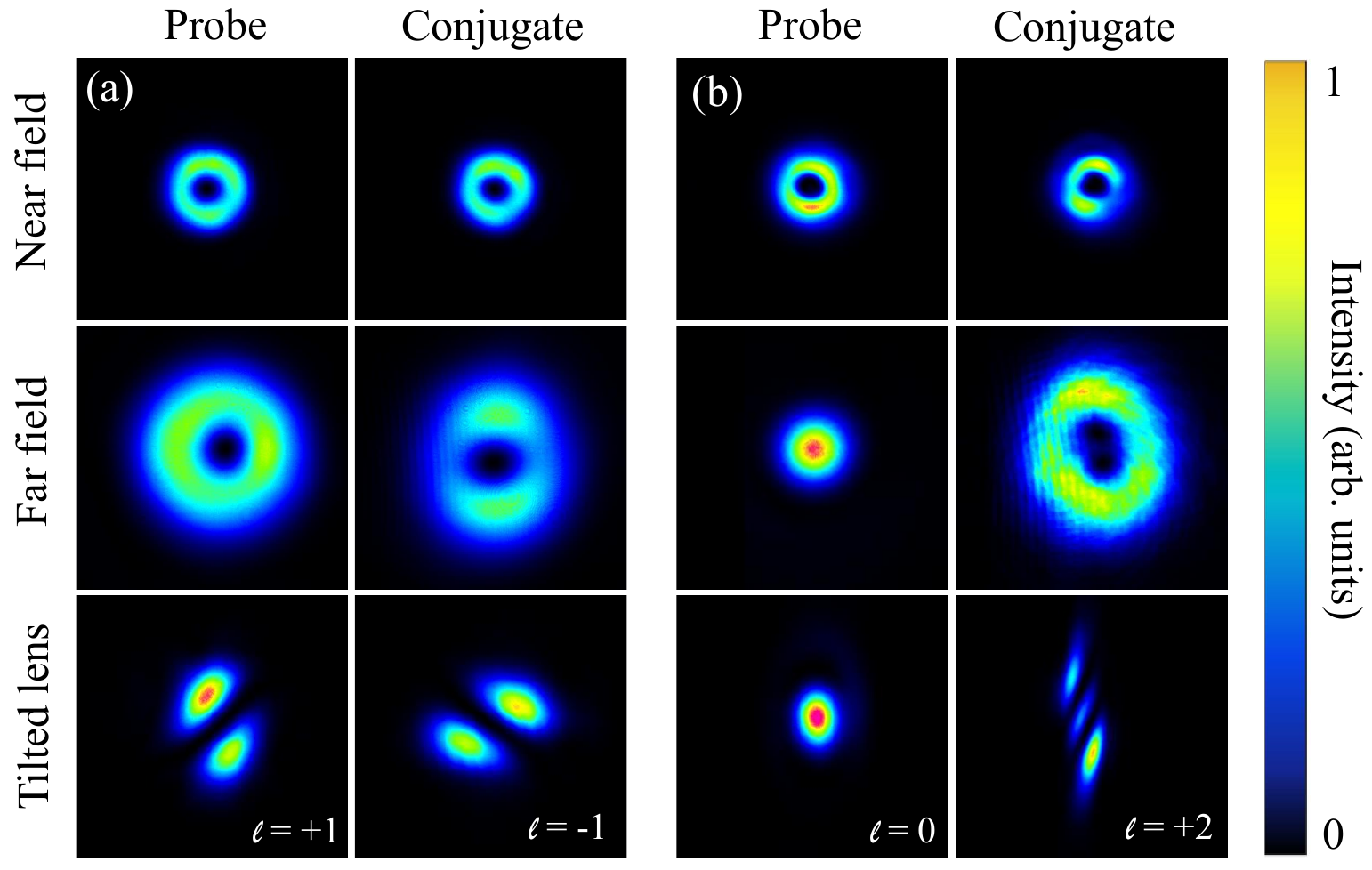} \caption{Experimental results for the transverse spatial intensity profiles of the probe and conjugate at the near and far field when (a) pump is Gaussian and seed probe is LG, and (b) pump is LG and seed probe is Gaussian. The third row shows the results with the tilted lens. The images are zoomed in for better visibility and are not to scale.}
    \label{fig:Fig6}
\end{center}
\end{figure}

Figure~\ref{fig:Fig6} shows the transverse spatial intensity profiles of the generated structured twin beams, corresponding to the later two input configurations (ii and iii). In Fig.~\ref{fig:Fig6}(a), the images of the twin beams are shown when the input pump is Gaussian and the seed probe is Laguerre Gaussian ($l$~=~+1). In the near field, both the probe and conjugate exhibit identical doughnut-shaped intensity profiles. In the far field, they retain the characteristic LG ring structure. To verify the OAM/topological charge carried by the beams, we use the tilted lens method \cite{Vaity}. As it can be seen from Fig.~\ref{fig:Fig6}(a) (last row), the probe carries the same topological charge as the input seed probe ($l$ = +1), while the conjugate beam carries a topological charge of -1, consistent with the OAM conservation law. 
\begin{figure}[t]
\begin{center}
    \includegraphics[width=0.45\textwidth]{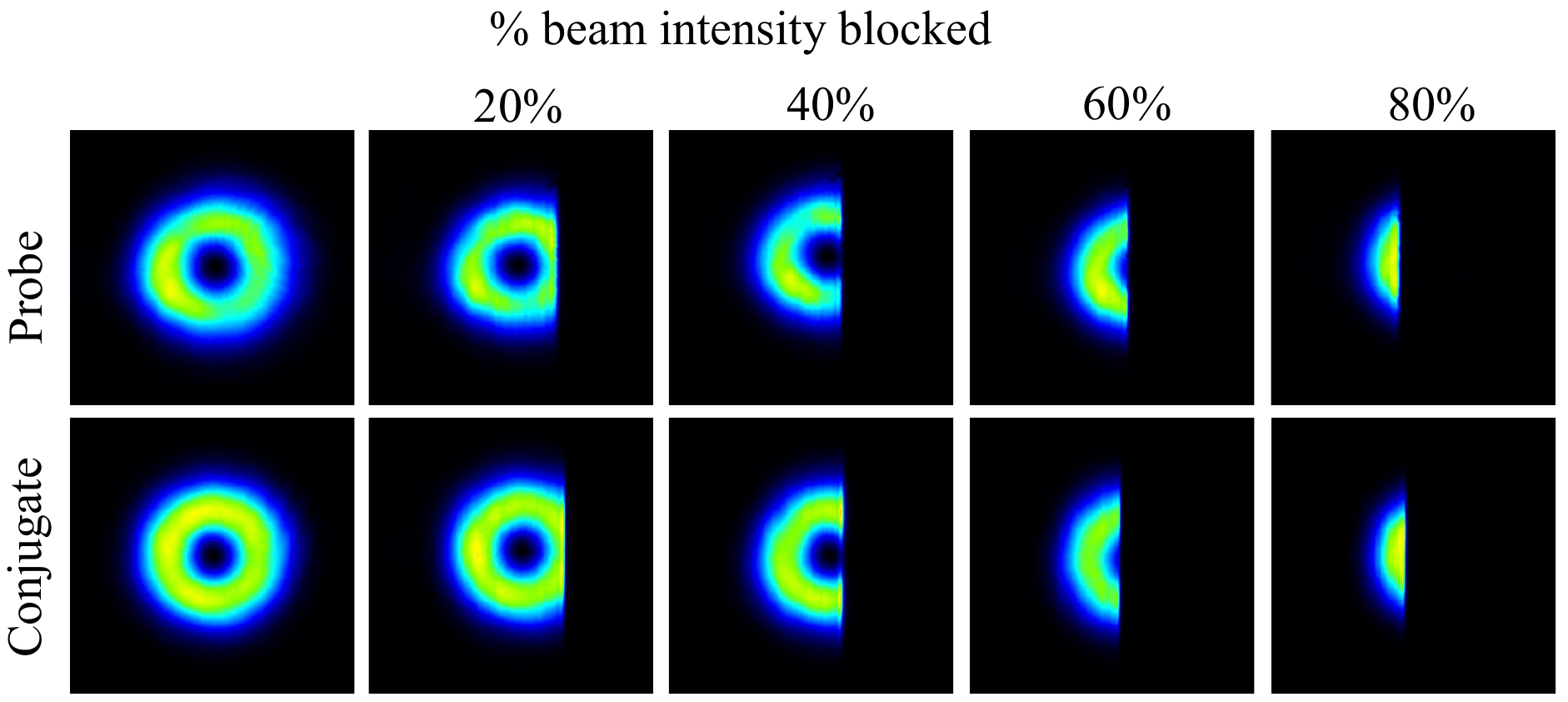} \caption{Spatial intensity profiles of generated probe and conjugate with different amounts of clipping.}
    \label{fig:Fig7}
\end{center}
\end{figure}

Figure~\ref{fig:Fig6}(b) shows the twin beam spatial profiles when the input pump is Laguerre Gaussian ($l$ = +1) and the seed probe is Gaussian. Interestingly, in the near field, both the probe and conjugate exhibit identical doughnut profiles, though the input seed probe is Gaussian. However, in the far field, the probe tends to become a Gaussian, while the conjugate retains the LG profile. We verified that the probe does not carry any topological charge while the conjugate carries a topological charge of +2 as seen in last row of Fig.~\ref{fig:Fig6}(b), obeying the OAM conservation law. The doughnut-shaped profile of the probe observed in the near field, despite carrying no OAM, is a direct consequence of spatially dependent electromagnetically induced transparency (EIT), as detailed in Ref~\cite{Jerin}. Note that the probe and conjugate always have identical spatial intensity profiles in the near field which is a consequence of localized emission of photon pair.

To investigate the spatial quantum correlation distribution between the twin beams, we select different spatial regions within the probe and conjugate beams and perform intensity difference noise measurement. An observation of the intensity difference noise below the shot noise limit (squeezing) confirms that the selected spatial subregions contain the quantum correlated pair of photons. Any presence of uncorrelated photons in the selected subregions introduces excess noise, thereby reducing the observed squeezing. This approach provides an insight into the distribution of transverse spatial quantum correlations between the twin beams. Here, we select the subregions in the twin beams by clipping the beams in varying amounts and measure the intensity difference noise with the selected regions at a fixed noise analysis frequency of 1.1 MHz. For each case, we compare the intensity difference noise of the beams with the shot noise limit at the same power. The two beams are clipped in equal amounts using knife edge in two different ways - from the same side of the beams (asymmetrically with respect to the pump) and from the opposite sides of the beams (symmetrically with respect to the pump). As an example, Fig.~\ref{fig:Fig7} shows the spatial profiles of the LG$_0^{+1}$ probe and LG$_0^{-1}$ conjugate beams with different amounts of beam intensity blocked from the same side.

\begin{figure*}[ht]
\begin{center}
    \includegraphics[width=1\textwidth]{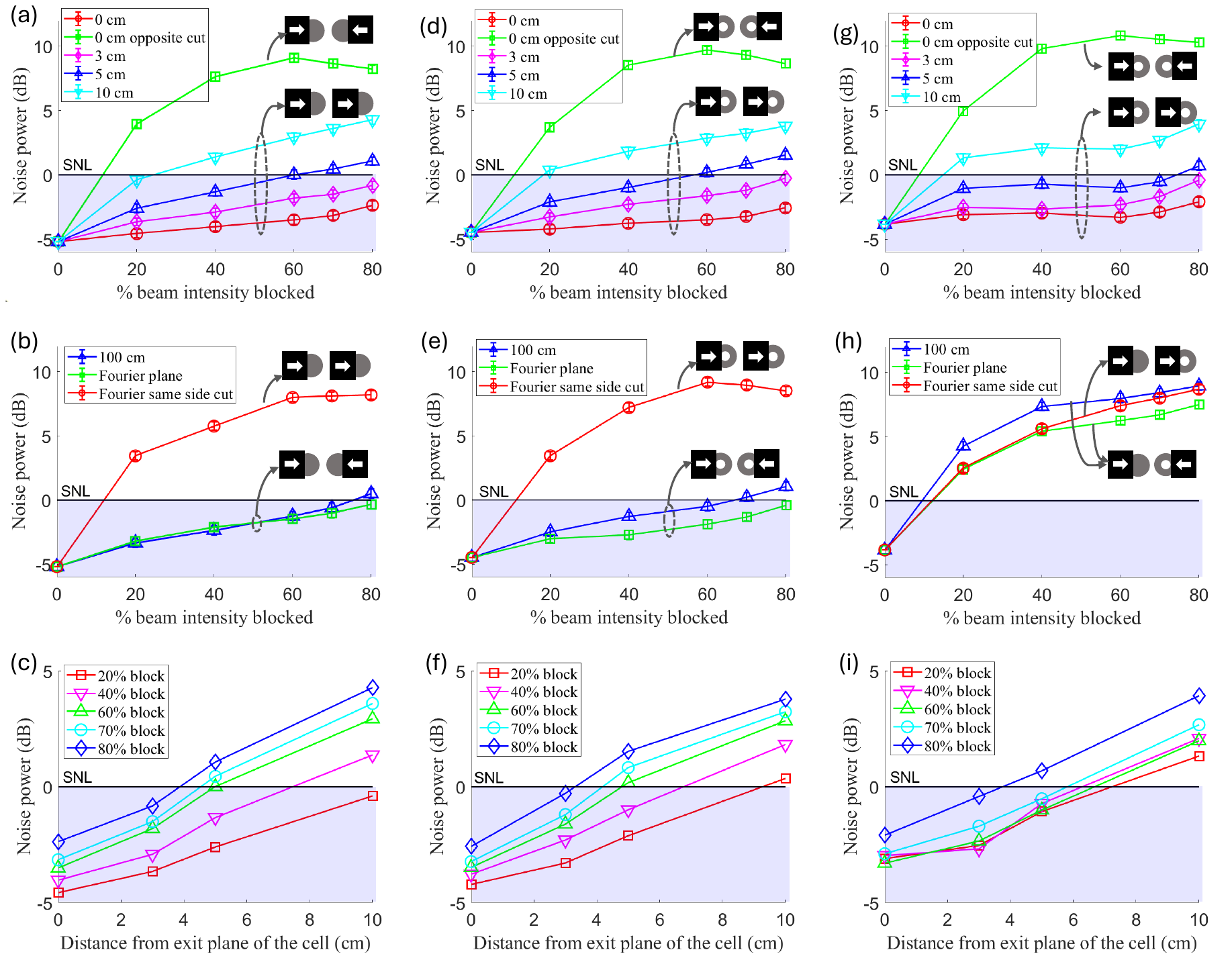} \caption{Normalized intensity difference noise power plotted as a function of the percentage of blocked beam intensity in both the probe and conjugate beams (first and second row), and propagation distance (third row), for three input configurations: (a, b, c) Gaussian pump with Gaussian probe, (d, e, f) Gaussian pump with LG probe, and (g, h, i) LG pump with Gaussian probe (lines connecting data points are shown to guide the eyes). The top row displays results in the near-field regime, while the middle row shows results in the far-field regime. The error bars plotted are for one standard deviation. The propagation distances of 0, 3, 5, 10, and 100 cm correspond to 0, 0.025, 0.042, 0.083, and 0.83 times the Rayleigh range of the pump ($z_{R}\sim120$~cm), respectively.}
    \label{fig:Fig8}
\end{center}
\end{figure*}

To investigate the propagation dynamics of spatial quantum correlations, the measurements are repeated at different propagation planes from the near-field to the far-field. The intensity difference noise normalized with respect to the shot noise is plotted against the percentage of the beam intensity blocked in Fig.~\ref{fig:Fig8} (top and middle row) at different propagation distances from the exit plane. The data points lying below the SNL (black trace) represent squeezing, which is a consequence of quantum correlations between the selected subregions of the twin beams. Additionally, we plot the intensity difference noise against the distance from the exit plane of the cell in Fig.~\ref{fig:Fig8} (last row) for different amounts of clipping.

In Fig.~\ref{fig:Fig8}(a), we present the results for the case when both the input pump and the seed probe are Gaussian. Without clipping the beams, we measured a maximum of 5.5 dB squeezing below the SNL. At the exit plane of the cell (z = 0 cm), we could observe significant amount of squeezing when equal portions of both beams are clipped from the same side (red trace, $\circ$). Moreover, squeezing  $\sim$2.5 dB is observed even when 80$\%$ of the beam intensity is blocked from the same side in both beams. This confirms the presence of strong position correlations in the near field, arising from the localized emission of photon pairs. As a result, both the correlated photons are detected at nearly the same position in the near field, and the extent of the spread in the correlated photon position is called coherence area. When 80$\%$ of both beams are blocked, the remaining 20$\%$ portions still contains many such coherence areas, resulting in significant amount of squeezing. In contrast, when the beams are clipped from opposite sides (green trace, \scalebox{0.7}{$\square$}), squeezing degrades rapidly due to the selection of uncorrelated regions, which introduces excess noise. At z = 3 cm (pink trace, $\diamond$) from the exit plane, the squeezing is observed up to 80$\%$ clipping, although the relative squeezing level has decreased compared to that at z = 0 cm. At z = 5 cm (blue trace, \scalebox{0.7}{$\Delta$}), the squeezing is further reduced and at z = 10 cm (cyan trace, \scalebox{0.7}{$\nabla$}), we observe almost no squeezing even for 20$\%$ clipping in both beams. This dynamics is more evident in Fig.~\ref{fig:Fig8}(c) where we plot intensity difference noise against the distance from the exit plane of the cell for different amount of beam clipping. The relative decrease in the squeezing level for a given amount of clipping at different distances from the exit plane can be attributed to both the increase in the coherence area along with the relative shift in its position with propagation, as evident from the vertical dashed lines in the last row of the theoretical plots of Figs.~\ref{fig:Fig2} and \ref{fig:Fig3}. Thus, in the selected areas of the twin beams, the probability of finding correlated photon pairs decreases as the beams propagate, leading to reduced squeezing.

On the other hand, in the far field (fourier plane of z = 0 cm, Fig.~\ref{fig:Fig8}(b)), the squeezing quickly degrades when both beams are clipped from the same side (red trace,~$\circ$). However, we observe squeezing when the two beams are clipped from opposite sides (green trace, \scalebox{0.7}{$\square$}), indicating strong position anti-correlations. In the far field also, squeezing is obtained with upto 80$\%$ of the beam intensity blocked. The strong position anti-correlations similar to the fourier plane were observed at around z = 100 cm ($\sim2z_R$, where $z_R$ is the Rayleigh length of the probe beam), as expected from the theory. 

Figures~\ref{fig:Fig8}(d, e, f) show the results for the configuration where the input pump is Gaussian, and the seed probe is LG. In this case, the spatial quantum correlation dynamics is similar to the case in Fig.~\ref{fig:Fig8}(a, b, c). This indicates that regardless of the probe spatial profile, the spatial quantum correlations are localized in the near field and far field when the pump is Gaussian. Additionally, the similarity in the squeezing traces across both input configurations suggests that the coherence area evolves similarly in both cases. This indicates that the seed probe primarily acts as a mode selector and does not change the size of the coherence area.  

In Figs.~\ref{fig:Fig8}(g, h, i), the results are shown for the configuration where the input pump is Laguerre-Gaussian and the seed probe is Gaussian. Here, without clipping the beams, we observed a maximum squeezing of 4 dB. In the near field (Fig.~\ref{fig:Fig8}(g)), similar to the case of the Gaussian pump, we observe localized position correlations, as is evident from the squeezing traces for the same side clipping (red trace,~$\circ$). The squeezing is lost quickly when the beams are clipped from the opposite side (green trace,~\scalebox{0.7}{\scalebox{0.7}{$\square$}}). The relative squeezing degrades with propagation as a result of the spread of the coherence area, and at z = 10 cm, the squeezing is not observed even with only 20$\%$ of the beams clipped, similar to the Gaussian pump case. However, here we observe nearly a flat trend in squeezing plot for different amount of clipping in the near field regime (z = 0, 3, 5 cm) compared to the case of Gaussian pump. This trend is also evident in Fig.~\ref{fig:Fig8}(i). This may be attributed to the asymmetry in the correlation pattern observed in Fig.~\ref{fig:Fig3}. In this regime, with the LG pump, most photon coincidence counts fall within a bright region whose width is smaller than that of the Gaussian case (this can be seen by comparing the width of the line profiles at z = 3, 5 and 10 cm for both input pump cases in theoretical plots of Figs.~\ref{fig:Fig2} and \ref{fig:Fig3}). This difference could influence the observed squeezing behavior due to the specific way in which the beams are clipped during measurement. In the Fourier plane (Fig.~\ref{fig:Fig8}(h)), the squeezing is immediately lost when both beams are clipped from the same (red,~$\circ$) or opposite side (green,~\scalebox{0.7}{$\square$}). This indicates that we cannot remove an independent mode from the conjugate which is correlated to the area removed from the probe, by clipping the beams in either of the ways. These observations are consistent with the delocalized correlations observed in the far-field theoretical results shown in Fig.~\ref{fig:Fig3}. 

Delocalized correlations are particularly useful in quantum information protocols, where they enable information encoding with complex quantum correlation distributions that spread across the transverse extent of individual beams, enhancing security in quantum cryptography \cite{Nirala, Verniere}. However, such correlations are less suited for quantum imaging, which relies on point-to-point spatial correlations to achieve high resolution. Nevertheless, delocalized correlations can be exploited for alternative strategies such as correlation-based pattern recognition \cite{Qiu}. 

\section{Conclusion} \label{Conclusion}

In this work, we have demonstrated both theoretically and experimentally how the spatial structure of the pump beam in a four-wave mixing process influences the evolution of spatial quantum correlations in bright twin beams. The theoretical analysis revealed that Gaussian and LG pumps produce fundamentally different correlation distributions across various propagation regimes. With a Gaussian pump, the spatial correlation pattern begins as a localized peak in the near field, transitions into a Gaussian shape in the intermediate regime, and retains this shape in the far field. In contrast, with an LG pump, the correlations start as localized, evolve into an asymmetric ring, and ultimately form a delocalized doughnut-shaped profile. Experimental measurements of spatial quantum correlations, based on intensity-difference noise analysis across different spatial subregions of the twin beams, showed good qualitative agreement with the theoretical predictions. These findings deepen our understanding of the spatial structure of quantum correlations. They also demonstrate how pump beam shaping can be harnessed to engineer nonclassical light fields for applications in quantum technologies, including quantum information processing and quantum imaging.
\appendix
\section*{Appendix}
\setcounter{equation}{0}
\renewcommand{\theequation}{A\arabic{equation}}

In the main text, we substitute $E(\rho,0)$ from Eq.~(8) in Eq.~(5) to obtain,  
\begin{align}
    \mathcal{E}(q_+) =\frac{1}{2 \pi} \frac{2}{\pi w^2 |l|!} \int \int d\phi d\rho & \left(\frac{2}{w^2}\right)^{|l|} \rho^{2|l| +1} e^{\frac{- 2\rho ^2}{w^2}} \nonumber \\ &e^{ -2il\phi }e^{-iq_+\rho \cos(\phi - \phi _+)},
\end{align}
where we have used the variable transformation as $q_+$~=~$q_{pr}$ + $q_c$,  $q_-$ = $q_{pr}$ - $q_c$ and $\phi_+$ = $\phi_{pr}$ + $\phi_c$. Now using the Jacobi-Anger expansion, 
\begin{align}
    e^{-iq_+\rho \cos(\phi - \phi _+)} = \sum_{n = -\infty}^{\infty} (-i)^n J_n(q_+\rho) e^{in(\phi - \phi _+)},
\end{align}
and using the fact that \begin{align}
    \int_0^{2\pi} d\phi \hspace{2pt}e^{i(n-m)\phi} = 2\pi\delta_{n,m},
\end{align}
the integral in Eq.~(A1) becomes,
\begin{align}
    \mathcal{E}(q_+) &= \frac{2}{\pi w^2 |l|!}\int  d\rho  \left(\frac{2}{w^2}\right)^{|l|} \rho^{2|l| +1} e^{\frac{- 2\rho ^2}{w^2}}J_{2l}(q_+\rho)e^{ -2il\phi_+ } \nonumber \\ &= \frac{1}{\pi|l|!} \frac{w^{2|l|}}{2^{3|l| + 1}} q_+^{2|l|} e^{\frac{- q_+ ^2  w^2}{8}}e^{ -2il\phi_+ },
\end{align}
which represents Eq.~(9) in the main text.

In order to obtain Eq.~(12) in the main text, we substitute Eq.~(10) in Eq.~(11),

\begin{align}
\mathcal{F}' = \frac{1}{(2\pi)^2}\frac{1}{\pi|l|!}\frac{w ^{2|l|}}{2^{3|l| + 1}} &\int \int  d\bm{q}_{pr} d\bm{q}_{c}  e^{-iq_{pr}  \cdot \rho_{pr}} e^{-iq_{c} \cdot \rho_{c}} q_+^{2|l|}  \nonumber \\ & \times e^{\frac{- q_+ ^2 w ^2}{8}} e^{ -2il\phi_+} \mbox{sinc}\left(\frac{L|q_-^2|}{8k_p}\right).
\end{align}

To simplify the integral, we perform a basis transformation
\begin{align}
    \mathcal{F}' = \frac{-1}{(2\pi)^2}\frac{1}{\pi|l|!}\frac{w ^{2|l|}}{2^{3|l| + 2}} &\int  d\bm{q}_{+}   e^{-i\frac{q_{+}  \cdot \rho_{+}}{2}}  q_+^{2|l|} e^{\frac{- q_+ ^2 w ^2}{8}}e^{ -2il\phi_+ } \nonumber \\ &\int d\bm{q}_{-} e^{-i\frac{q_{-} \cdot \rho_{-}}{2}} \mbox{sinc}\left(\frac{L|q_-^2|}{8k_p}\right).
\end{align}

We now solve the $q_+$ and $q_-$ part separately by defining
 \begin{align}
     I_{q_+} &= \int  d\bm{q}_{+}   e^{-i\frac{q_{+}  \cdot \rho_{+}}{2}}  q_+^{2|l|} e^{\frac{- q_+ ^2 w ^2}{8}}e^{ -2il\phi_+ } \nonumber \\ &= \int \int d\phi_+  dq_{+}   e^{-i\frac{q_{+}\rho_{+}}{2} \cos(\phi_+ - \theta_+)}  q_+^{2|l|+1} e^{\frac{- q_+ ^2 \omega ^2}{8}}e^{ -2il\phi_+ } \nonumber \\ &= 2\pi\int  dq_{+}   q_+^{2|l| + 1} e^{\frac{- q_+ ^2 w ^2}{8}} J_{2l}(\frac{q_+\rho_+}{2})e^{ -2il\theta_+ }.
 \end{align}
 
Now the other integral involving $q_-$ part, 
\begin{align}
    I_{q_-} &= \int d\phi_- dq_{-} q_- e^{-i\frac{q_{-} \cdot \rho_{-}}{2}} \mbox{sinc}\left(\frac{L|q_-^2|}{8k_p}\right) \nonumber \\ &= 2\pi\int  dq_{-} q_-  \mbox{sinc}\left(\frac{L|q_-^2|}{8k_p}\right)J_0(\frac{q_-\rho_-}{2}).
\end{align}

Thus, the two-photon amplitude in the position basis becomes,
\begin{align}
    \mathcal{F}' = - \frac{1}{\pi|l|!}\frac{w ^{2|l|}}{2^{3|l| + 2}} \int\int  dq_{-}   dq_{+} q_-  q_+^{2|l| + 1} e^{\frac{- q_+ ^2 w^2}{8}} e^{ -2il\theta_+ } \nonumber \\  \mbox{sinc}\left(\frac{L|q_-^2|}{8k_p}\right)J_{2l}(\frac{q_+\rho_+}{2})J_0(\frac{q_-\rho_-}{2}).
\end{align}

\begin{widetext}

In the main text, to obtain the propagated two-photon probability amplitude at a distance $z$, Eq.~(12) and Eq.~(13) are substituted in Eq.~(14) as,

\begin{align}
\tilde{\mathbb{F}} = \frac{C}{z^2}\left(\frac{k_p}{2\pi}\right)^2 e^{2ik_pz}e^{i\frac{k_p}{2z}(\rho_{pr}^{'2}+\rho_{c}^{'2})}\int \int \int\int  dq_{-}  dq_{+} d\bm{\rho}_{pr} d\bm{\rho}_{c} q_-  q_+^{2|l| + 1} e^{\frac{- q_+ ^2 w ^2}{8}} e^{ -2il\theta_+ }  &\mbox{sinc}\left(\frac{L|q_-^2|}{8k_p}\right)J_{2l}(\frac{q_+\rho_+}{2})J_0(\frac{q_-\rho_-}{2}) \nonumber \\ & \times e^{i\frac{k_p}{2z}(\rho_{pr}^{2}+\rho_{c}^{2})}e^{-i\frac{k_p}{z}(\rho_{pr}^{'}\cdot \rho_{pr} + \rho_{c}^{'}\cdot \rho_{c})},
\end{align}
where $C = \frac{1}{\pi|l|!}\frac{w ^{2|l|}}{2^{3|l| + 2}}$. Now we change the variables $\bm{\rho}_+ = \bm{\rho}_{pr} + \bm{\rho}_c$, $\bm{\rho_-} = \bm{\rho}_{pr} - \bm{\rho_c}$, and $\bm{\rho_+^{'}} = \bm{\rho}_{pr}^{'} + \bm{\rho}_c^{'}$, $\bm{\rho}_-^{'} = \bm{\rho}_{pr}^{'} - \bm{\rho}_c^{'}$ and we obtain,

\begin{align}
	\tilde{\mathbb{F}} = \frac{C}{z^2}\left(\frac{k_p}{2\pi}\right)^2e^{2ik_pz}e^{i\frac{k_p}{4z}(\rho_{+}^{'2}+\rho_{-}^{'2})} \int\int  dq_{-}  & dq_{+}  q_-  q_+^{2|l| + 1} e^{\frac{- q_+ ^2 \omega ^2}{8}}  \mbox{sinc}\left(\frac{L|q_-^2|}{8k_p}\right) \nonumber \\ & \times \int \int d\bm{\rho}_{+} d\bm{\rho}_{-}  J_{2l}(\frac{q_+\rho_+}{2})J_0(\frac{q_-\rho_-}{2})e^{i\frac{k_p}{4z}(\rho_{+}^{2}+\rho_{-}^{2})}e^{-i\frac{k_p(\rho_{+}^{'}\cdot \rho_{+} + \rho_{-}^{'}\cdot \rho_{-})}{2z}}e^{ -2il\theta_+}.
\end{align}

After solving the azimuthal part, the integral reduces to,
\begin{align}
	\tilde{\mathbb{F}} = \frac{Ck_p^2}{z^2}e^{2ik_pz}e^{i\frac{k_p}{4z}(\rho_{+}^{'2}+\rho_{-}^{'2})}e^{ -2il\theta^{'}_+ } & \int\int  dq_{-} dq_{+}  q_-  q_+^{2|l| + 1} e^{\frac{- q_+ ^2 \omega ^2}{8}}  \mbox{sinc}\left(\frac{L|q_-^2|}{8k_p}\right) \nonumber \\ & \times \int  d\rho_{+} \rho_+  J_{2l}(\frac{q_+\rho_+}{2})J_{2l}(\frac{k_p\rho^{'}_+\rho_+}{2z})e^{i\frac{k_p\rho_{+}^{2}}{4z}} \int d\rho_{-} \rho_{-} J_0(\frac{q_-\rho_-}{2})J_{0}(\frac{k_p\rho^{'}_-\rho_-}{2z})e^{i\frac{k_p\rho_{-}^{2}}{4z}}.
\end{align}
 Now we introduce a small Gaussian decay factor $\left(\lim_{\epsilon \to 0^+}e^{-\epsilon\rho_+^2}\right)$ to turn the integrals into standard integrals as,
\begin{align}
	\tilde{\mathbb{F}} = lim_{\epsilon \to 0}\frac{Ck_p^2}{z^2} e^{ -2il\theta^{'}_+ }e^{2ik_pz} & e^{i\frac{k_p}{4z}(\rho_{+}^{'2}+\rho_{-}^{'2})} \int\int  dq_{-}   dq_{+}  q_-  q_+^{2|l| + 1} e^{\frac{- q_+ ^2 \omega ^2}{8}}  \mbox{sinc}\left(\frac{L|q_-^2|}{8k_p}\right)  \nonumber \\ & \times \int  d\rho_{+}  \rho_+ J_{2l}(\frac{q_+\rho_+}{2})J_{2l}(\frac{k_p\rho^{'}_+\rho_+}{2z})e^{-(\epsilon - i\frac{k_p}{4z})\rho_{+}^{2}} \int d\rho_{-} \rho_{-} J_0(\frac{q_-\rho_-}{2})J_{0}(\frac{k_p\rho^{'}_-\rho_-}{2z})e^{-(\epsilon - i\frac{k_p}{4z})\rho_{-}^{2}}.
\end{align}
Now we use the standard integral
\begin{align}
    \int_0^{\infty} te^{-p^2 t^2}J_{\nu}(at)J_{\nu}(bt) dt = \frac{e^{\frac{-(a^2 + b^2)}{4p^2}}}{2p^2}I_{\nu}(\frac{ab}{2p^2}),
\end{align}
where I is the modified Bessel function. With this, we obtain, 

\begin{align}
\tilde{\mathbb{F}} = 4Ce^{ -2il\theta^{'}_+ } e^{2ik_pz} \int   dq_{+}    q_+^{2|l| + 1} e^{-(\frac{\omega ^2}{8} + \frac{iz}{4k_p})q_+^2}   I_{2l}(\frac{iq_+ \rho'_+}{2})  \int  dq_{-} q_-  e^{-i\frac{z q_-^2}{4k_p}} \mbox{sinc}\left(\frac{L|q_-^2|}{8k_p}\right) I_{0}(\frac{iq_- \rho'_-}{2}). 
\end{align}
The $q_+$ integral takes the form of a standard integral
\begin{align}
   \int dt t^{\nu +1} I_{\nu}(bt)e^{-p^2t^2} = \frac{b^{\nu}}{(2p^2)^{\nu +1}}e^{\frac{b^2}{4p^2}}. 
\end{align}
Thus we finally get 
\begin{align}
\tilde{\mathbb{F}} = \frac{1}{2^{2|l|+1}}\frac{4C}{(\frac{\omega ^2}{4} + \frac{iz}{2k_p})^{2l+1}}e^{2ik_pz}e^{ -2il\theta^{'}_+ } (\rho^{'}_+)^{2|l|} e^{-\frac{\rho^{'2}_+}{16(\frac{\omega ^2}{8} + \frac{iz}{4k_p})}} \int  dq_{-} q_-  e^{-i\frac{z q_-^2}{4k_p}} \mbox{sinc}\left(\frac{L|q_-^2|}{8k_p}\right) I_{0}(\frac{iq_- \rho' _-}{2}) .
\end{align}

To further simplify our calculations, we approximate the sinc term using a Cosine-Gaussian approximation sinc($x^2$) = cos$(ax^2)$$e^{-b x^2}$ where $a$ and $b$ are constant factors. On solving, we obtain the final analytical expression for the propagated two-photon amplitude as,

\begin{align}
\tilde{\mathbb{F}} = \frac{C'}{(\frac{w ^2}{4} + \frac{iz}{2k_p})^{2|l|+1}} (\rho^{'}_+)^{2|l|}e^{ -2il\theta^{'}_+ }e^{2ik_pz} e^{-\frac{\rho^{'2}_+}{16(\frac{w ^2}{8} + \frac{iz}{4k_p})}} &\left[ \frac{1}{2\left( b\frac{L}{8k_p} + i\left(\frac{z }{4k_p} + a \frac{L}{8k_p}\right)\right)}e^{\frac{-\rho _{-}^{'2}}{16\left(b \frac{L}{8k_p} + i\left(\frac{z }{4k_p} + a \frac{L}{8k_p}\right)\right)}}\right. \nonumber \\ &\left. +\frac{1}{2\left(b \frac{L}{8k_p} + i\left(\frac{z}{4k_p} - a \frac{L}{8k_p}\right)\right)}e^{\frac{-\rho _{-}^{'2}}{16\left(b \frac{L}{8k_p} + i\left(\frac{z }{4k_p} - a \frac{L}{8k_p}\right)\right)}}\right],
\end{align}
where $C' = \frac{1}{\pi |l|!}\frac{w ^{2|l|}}{2^{5|l| + 1}}$ is a constant. The two-photon probability distribution is obtained by taking $|\tilde{\mathbb{F}}|^2$.

\end{widetext}

\end{document}